\documentclass[11pt,aip,jcp,preprint,floatfix]{revtex4-1}
\usepackage{graphicx}
\usepackage{dcolumn}
\usepackage{latexsym}
\usepackage{amsmath}
\usepackage{amssymb}
\usepackage{color}
\begin{document}
\title{Dielectric Spectroscopy and Ultrasonic Study of Propylene Carbonate under Ultra-high Pressures.}
\author{M.V.Kondrin}
\email{mkondrin@hppi.troitsk.ru}
\author{E.L. Gromnitskaya}
\affiliation{Institute for High Pressure Physics RAS, 142190 Troitsk, Moscow region, Russia}
\author{A.A.Pronin}
\affiliation{General Physics Institute RAS, 117942 Moscow, Russia}
\author{A.G.Lyapin}
\author{V.V. Brazhkin}
\affiliation{Institute for High Pressure Physics RAS, 142190 Troitsk, Moscow region, Russia}
\author{A.A.Volkov}
\affiliation{General Physics Institute RAS, 117942 Moscow, Russia}
\begin{abstract}
We present the high pressure dielectric spectroscopy (up to 4.2 GPa) and ultrasonic study (up to 1.7 GPa) of liquid and glassy propylene carbonate (PC). Both of the methods provide complementary pictures of the glass transition in PC under pressure. No other relaxation processes except $\alpha$-relaxation have been found in the studied pressure interval. The propylene carbonate liquid is a glassformer where simple relaxation and the absence of $\beta$-relaxation are registered  in the record-breaking ranges of pressures and densities. The equation of state of liquid PC was extended up to 1 GPa from ultrasonic measurements of bulk modulus and is in good accordance with the previous equations developed from volumetric data. We measured the bulk and shear moduli and Poisson's ratio of glassy PC up to 1.7 GPa. Many relaxation and elastic properties of PC can be qualitatively described by the soft-sphere or Lennard-Jones model. However, for the quantitative description of entire set of the experimental data, these models are insufficient. Moreover,  the Poisson coefficient value for glassy PC indicates a significant contribution of non-central forces to the intermolecular potential. The well-known correlation between Poisson's ratio and fragility index (obtained from dielectric relaxation) is confirmed for PC at ambient pressure, but it is violated with pressure increase. This indicates that different features of the potential energy landscape are responsible for the evolution of dielectric response and elasticity with pressure increase.
\end{abstract}
\pacs{05.70.-a,62.10.+s,62.20.-x,62.50.-p,64.70.kj}
\maketitle
\section{Introduction}
Propylene carbonate (PC) is a well-known molecular glassformer which was quite extensively studied previously \cite{schneider:pre98,wang-velikov-angell:jcp02}. Upon cooling down to 158 K at moderate rates it easily vitrificates. Interesting to note, that its neighbourgh in the homological sequence, namely  ethylene carbonate, having practically the same dipole moment but a more symmetrical molecule (without the protruding methyl group) is a crystal at room temperature \cite{brown:ac54}. On the other hand, PC at ambient condition is  a van-der-Waals liquid and an aprotic solvent (and in this quality it is widely used in practice); therefore, the intermolecular interaction in this liquid is considerably simpler than in hydrogen-bonded ones. Due to its simple structure (propylene carbonate is a carbonated version of propylene glycole, see inset in Fig.~\ref{fig:cc}), a relatively compact molecule and the absence of hydrogen bonds, propylene carbonate may provide a suitable object for both theoretical and experimental study of correlation between the intermolecular bonding and the profile of liquid energy landscape, on the one hand, and the relaxation dynamics leading to vitrification of the liquid, on the other \cite{stillinger:n01}. 

By using the high pressure technique this study can be made more informative because the energy landscape modification in the pressure range up to 10 GPa  should be quite significant \cite{voigtmann:prl08}. So the observation of possible changes in the relaxation dynamics upon the high pressure application may give valuable information about the character of intermolecular forces  and the relaxation kinetics in the liquid.

Structural relaxation in liquids manifests itself in the dielectric spectroscopy \cite{davidson:jcp51} and ultrasound velocity \cite{litovitz:jasa58,jeong-yoon:pra86,jeong-yoon:pra87} measurements  as a broad dispersion feature in the frequency range up to several Thz. Upon supercooling and  subsequent vitrification, the characteristic frequency of the structural relaxation becomes considerably lower and shifts toward the radio-frequency and even sub-Hertz range. There is a number of works where the pressure dependence of structural relaxation in PC was studied by dielectric spectroscopy methods \cite{pawlus:prb04,bielowka:pre04,reiser-kasper:prb05,reiser-kasper:epl06,rzoska:jpcm08} but the data on  ultrasound velocity measurements are still absent, although this method not only provides  the data on the relaxation characteristics but also sheds the light upon the thermodynamic data, such as the equation of state for the liquid. In the pressure range up to 1.6 GPa, the authors of the Refs.~\onlinecite{bielowka:pre04,casalini:jcp08,reiser-kasper:prb05,rzoska:jpcm08} agree that the structural relaxation in PC displays a quite simple scaling dependence which can be understood in terms of the soft-spheres model describing intermolecular forces as effectively repulsive with the power dependence on the distance $r$ between molecules:
\begin{equation}
U_{ss}(r)=A\left(\frac{r}{r_0}\right)^{-n}
\label{eq:ss}
\end{equation}
This potential can be viewed as an abbreviated version of the $m-n$ potential
\begin{equation}
U_{LJ}(r)=A\left(\frac{r}{r_0}\right)^{-n}-B\left(\frac{r}{r_0}\right)^{-m} 
\label{eq:lj}
\end{equation}
, which is a more general version of the Lennard-Jones one ($m=6$ and $n=12$). The latter may be a good approximation of interactions that take place in liquid PC, since this potential is quite typical of molecular systems.

 Thermodynamic properties of the soft-sphere model can be derived using the Klein theorem \cite{berlin:jcp52,hoover:jcp71,en*stishov:up74}, but the solution for the kinetic properties will be more complicated. Taking into account a direct analogy \cite{pawlus:prb04,casalini-roland:prb05} sometimes supported by deeper theoretical consideration  \cite{hoover:jcp71,en*stishov:up74,en*zhakhovsky:zhetp94,fomin:jetpl12}, one may write a scaling relationship for the characteristic  time  of the structural relaxation (also known as $\alpha$-relaxation) $\tau$ in the form:
\begin{equation}
\tau=f\left(TV^{n/3}\right)
\label{eq:scaling}
\end{equation}
To validate this relation, one needs to know the equation of state of liquid PC. In the low-pressure region below 1 GPa, there are two independent approximations of the equation of state with the Tait formula \cite{bielowka:pre04,reiser-kasper:prb05} which enable the number of authors to corroborate the scaling relation for PC. The obtained values of exponent $n$ slightly differ: dielectric measurements yield the values  $n=11.1$ \cite{bielowka:pre04} and $n=12.6$ \cite{reiser-kasper:prb05} while viscosity measurements -- $n=12.9$ \cite{casalini:jcp08}. Nevertheless these values are remarkably close to the number 12 which makes one recall the well-known ``12-6'' Lennard-Jones potential. Strictly speaking, this exponent value for the repulsive part of intermolecular potential is not well grounded theoretically but it is often recorded in practice, even in very simple cases, such as the noble gases, where it can be deduced from the shape of the melting curve \cite{stishov:jetp06}.

Despite impressive agreement between the soft-sphere model predictions and  the experimental results obtained in Refs.~\onlinecite{bielowka:pre04,reiser-kasper:prb05} we are not satisfied with it because these conclusions were drawn from the experimental data consisting of a few isotherm measurements while the isobaric data are available only for a limited pressure range below 0.6 GPa \cite{reiser-kasper:epl06}. Moreover, the recent measurements of structural relaxation under pressures up to 6.5 GPa in another well-known small-molecule glassformer glycerol \cite{pronin:pre10,pronin:jetpl10} reveal that in the supercooled liquid  upon rising the pressure over 2 GPa qualitative change of dynamics takes place which splits  the main dispersive curve into two sub-processes (the high-frequency one is also known as the secondary or $\beta-$ or Johari-Goldstein relaxation). This crossover was earlier observed in many molecular glassformers \cite{paluch:prl03} but for a long time glycerol was an exception due to a relatively high pressure where this crossover took place. Interesting enough that the earlier experiments on the aging of glassformers \cite{schneider:prl00,ngai:jcp01} suggest that a similar type of high pressure relaxation exists for  PC as well. However, as it will be shown below in this paper the behavior of structural relaxation in PC and glycerol in the high pressure region turns out to be  quite different.

In addition to dielectric spectroscopy, many other experimental techniques, including the Brillouin spectroscopy and other light-scattering methods \cite{dreyfus:prl92,elmroth:prl92,tabellout:prb95,shen:pre00}, viscosity\cite{casalini:jcp08,herbst:n93,hiki:msea04} and internal friction \cite{hiki:msea04} measurements, ultrasonic \cite{litovitz:jasa58,jeong-yoon:pra86,jeong-yoon:pra87,dreyfus:prl92,tabellout:prb95,shen:pre00,hiki:msea04,cutroni:jcp01,carini:jpcm06} and mechanical dynamic methods \cite{schroter:jncs02,tabellout:prb95,mandanici:jcp05}, neutron scattering \cite{shen:pre00,toulouse:pa93}, NMR \cite{chang:jpcb97}, calorimetric \cite{carini:jpcm06,ngai:prb90} and thermal expansion \cite{carini:jpcm06} techniques, can be applied to study the glass transition. Combination of different techniques provides, as a rule, more detailed and complete picture of relaxation processes in supercooled liquids and glasses \cite{dreyfus:prl92,tabellout:prb95,shen:pre00,hiki:msea04,carini:jpcm06,ngai:prb90}. The key problems in comparative studies of supercooled liquids (by different techniques) usually include a relation between rotational and translational self-diffusion and molecular dynamics \cite{shen:pre00,chang:jpcb97,cicerone:jcp95,hall:prl97}, as well as validation of the Stokes-Einstein and Debye-Stokes-Einstein equations \cite{casalini:jcp08,fomin:jetpl12,chang:jpcb97,cicerone:jcp95} near the glass transition temperature $T_g$. For molecular glassformers, the orientational molecular dynamics (i.e., rotational molecular movements) can be tested by dielectric spectroscopy and depolarized light-scattering spectroscopy \cite{shen:pre00}, while other methods are much more sensitive to translational molecular movements. Ultrasonic technique \cite{litovitz:jasa58,
jeong-yoon:pra86,jeong-yoon:pra87,dreyfus:prl92,tabellout:prb95,
shen:pre00,hiki:msea04,cutroni:jcp01,carini:jpcm06} is an example of the methods of the last type, in which the sound attenuation provides information on the relaxation properties, while wave velocities -- on the elasticity and subsequently equation of state. The ultrasonic study of elastic properties also enables one to evaluate effective parameters of intermolecular forces.

As far as we know, there were only a few attempts to study the glass transition by ultrasonic techniques under pressure (e.g., in glycerol \cite{slie:jcp68}) in contrast to the high-pressure applications of dielectric spectroscopy (see \cite{pawlus:prb04,bielowka:pre04,reiser-kasper:prb05,reiser-kasper:epl06,rzoska:jpcm08,casalini-roland:prb05} and references therein) and viscosity measurements (e.g., Refs. \onlinecite{casalini:jcp08,herbst:n93} and references therein).

Here we combine the dielectric spectroscopy and pulse ultrasonic method and apply them to propylene carbonate under ultra-high pressure. The rotational molecular movement is dominating in the dielectric response, while attenuation of ultrasound waves highlights the contribution of translational (basically transverse) modes. Thus, the comparison of both methods is actual to test PC as a substance well described by the soft-sphere model \cite{fomin:jetpl12}. On the other hand, the pressure dependence of elastic characteristics (elastic moduli and Poisson's ratio) obtained by ultrasonic method gives an  opportunity to correlate the elasticity and relaxation properties of the PC glass, e.g., fragility. Another motivation for using the  ultrasonic method is a possibility to check central or non-central character of intermolecular forces through determination of the Poisson coefficient. 

\section{Experimental setup}
\label{sec:meth}
The samples of 99.5 \% pure propylene carbonate were obtained from Acros Organics and were used as delivered. High-pressure dielectric spectroscopy experiments were carried out in Toroid type anvils  \cite{hpr:khvostantsev2004} in the P-T region P $<$ 4.1 GPa and 210 K $<$ T $<$ 410 K. As a sensor we  used the plane capacitor made of copper plates separated by fixed size Teflon spacers with 1.5 mm diameter holes (having typical empty capacitance about 10~pF) immersed into PC contained in a Teflon cell. Therefore, the studied liquid also served as a pressure-transmitting medium. The signal from the capacitor was registered with Quadtech-7600 LCR-meter in the frequency range 10 Hz -- 2 Mhz. The pressure and temperature sensors consisting of manganin wire and chromel-alumel thermocouple were also put into the Teflon cell. The overall precision of external parameters determination were better than 0.05 GPa and 0.1 K. In every experimental run we swept in a stepwise manner 2 or 3 isobars and isotherms (as schematically shown by arrows in Fig.~\ref{fig:p-t}). That is, we first raised the pressure until the frequency of $\alpha$-relaxation became lower than minimum of our frequency range, whereupon we started to heat our setup until this frequency became greater than maximal frequency of the range. In the course of the experiment, we took special care that the sample was not put into the vitreous state (which could be easily checked by the breakage of the manganin sensor). The speed of temperature and pressure changes was not strictly controlled but the typical rates can be estimated as 0.02 GPa/min and 0.05 K/min respectively. The dielectric spectroscopy data presented in this paper  were collected during several runs of such a type. 

The ultrasonic study of elastic characteristics was carried out with a high-pressure ultrasonic piezometer based on the piston-cylinder device \cite{stalgorova:iet96} with the registration system of the transmitted and reflected sound constructed on the basis of the National Instruments PXI platform. Changes in ultrasonic signal paths were measured with 0.005 mm uncertainty using dial-type micrometer indicators. The sample volume under pressure was initially determined from the change in the length, and subsequently the equation of state above glass transition was calculated by the recursive (with respect to density) integration of compressibility (reverse bulk modulus) with accuracy of several \%. The changes in the transit times (to $\pm 0.001$ $\mu$s) of longitudinal and transverse ultrasonic pulses were measured by the pulsed method using x-cut and y-cut quartz or LiNbO$_3$ plates as piezoelectric sensors with carrier frequencies equal to 4.7 MHz for the  transverse wave and  10 MHz for the longitudinal one. The transit time was measured not only by time shift of the passed sound wave \cite{stalgorova:iet96} but also by the cross-correlation and autocorrelation method \cite{pantea:rsi05}, when one could determine the absolute transit time of the sound wave in a sample. The estimated pressure uncertainty was $\approx$0.03 GPa. Corrections for the chamber deformation with the  pressure and temperature  variations were determined in separate  experiments. This method of ultrasonic measurement of liquids with the high-pressure piezometer was previously developed for liquid methanol \cite{gromnitskaya:jetpl04} and gallium \cite{grom:prl07,lyapin:jetp08}, using original copper-teflon capsules. The stability of these capsules against the liquid leakage restricted the maximum attainable pressure for measurements by the value about 1 GPa. The main source of uncertainty for measurements of the ultrasonic wave velocity in liquids is associated with  the determination of the sample height (i.e., the path length for the wave). We estimated the liquid PC sample height from the sample mass and determined it more accurately from the literature data on the wave velocity at ambient pressure. The details of this procedure will be presented later together with the experimental results. Glassy PC was studied up to 1.7 GPa. The temperature in the working volume was measured with four copper-constantan thermocouples located in the immediate vicinity of the sample; the temperature gradient across the sample did not exceed 1--2 K. In order to calculate adiabatic  shear G and bulk B moduli, we used the approximation of a homogeneous isotropic medium: 
\begin{equation}
\begin{split}
G&=\rho v_t^2 \\
B+{4\over 3}G&=\rho v_l^2
\end{split}
\label{eq:B-G}
\end{equation}
where $\rho$ is the density, $v_t$ and $v_l$ are the transverse and longitudinal ultrasonic velocities, respectively. Poisson's coefficient (also known as the Poisson coefficient) which quantifies transverse response to uniaxial stress was also calculated in the approximation of an isotropic medium according to the equation: 
\begin{equation}
\sigma={(3B-2G)\over (6B+2G)}
\label{eq:sigma}
\end{equation}
Intuitively it is clear that Poisson's ratio is correlated with the orientation of intermolecular forces. Indeed, solid systems with only central forces should have Poisson's ratio equal or close to 0.25. Specifically, for simple crystalline lattices with central forces it can be  easily shown that the Cauchy relations produce the relation for the Voigt shear modulus $G=(3/5)B$ ($P=0$) \cite{br:jetpl01}, so  Eq.~(\ref{eq:sigma}) yields $\sigma=0.25$. This value divides, for example, covalent substances ($\sigma <0.25$) with a high shear elasticity from plastic metals ($\sigma \sim 0.3$--0.4) \cite{br:pma02}.

The study of ultrasound wave attenuation plays the key role in the ultrasonic testing of the relaxation properties of a supercooled liquid \cite{jeong-yoon:pra86,jeong-yoon:pra87,cutroni:jcp01}. Accurate measurement of attenuation coefficient in the facility with the long input acoustic lines \cite{stalgorova:iet96} demands a comparison of the amplitudes of waves passed through the sample by one and two or three (the more is the better) times. Unfortunately, such a type of measurements was impossible in the regime of strong attenuation near the glass transition temperature $T_{\rm g}$. Moreover, a stray ultrasound pulse propagating through the cylinder of high pressure ultrasonic piezometer makes impossible any accurate measurements in the regime of strong attenuation in the sample. For this reason the attenuation coefficient was estimated here by the amplitude of the wave after its single pass through the piezometer with the sample.
\section{Dielectric spectroscopy data}
\begin{figure}
  \includegraphics[width=0.8\columnwidth]{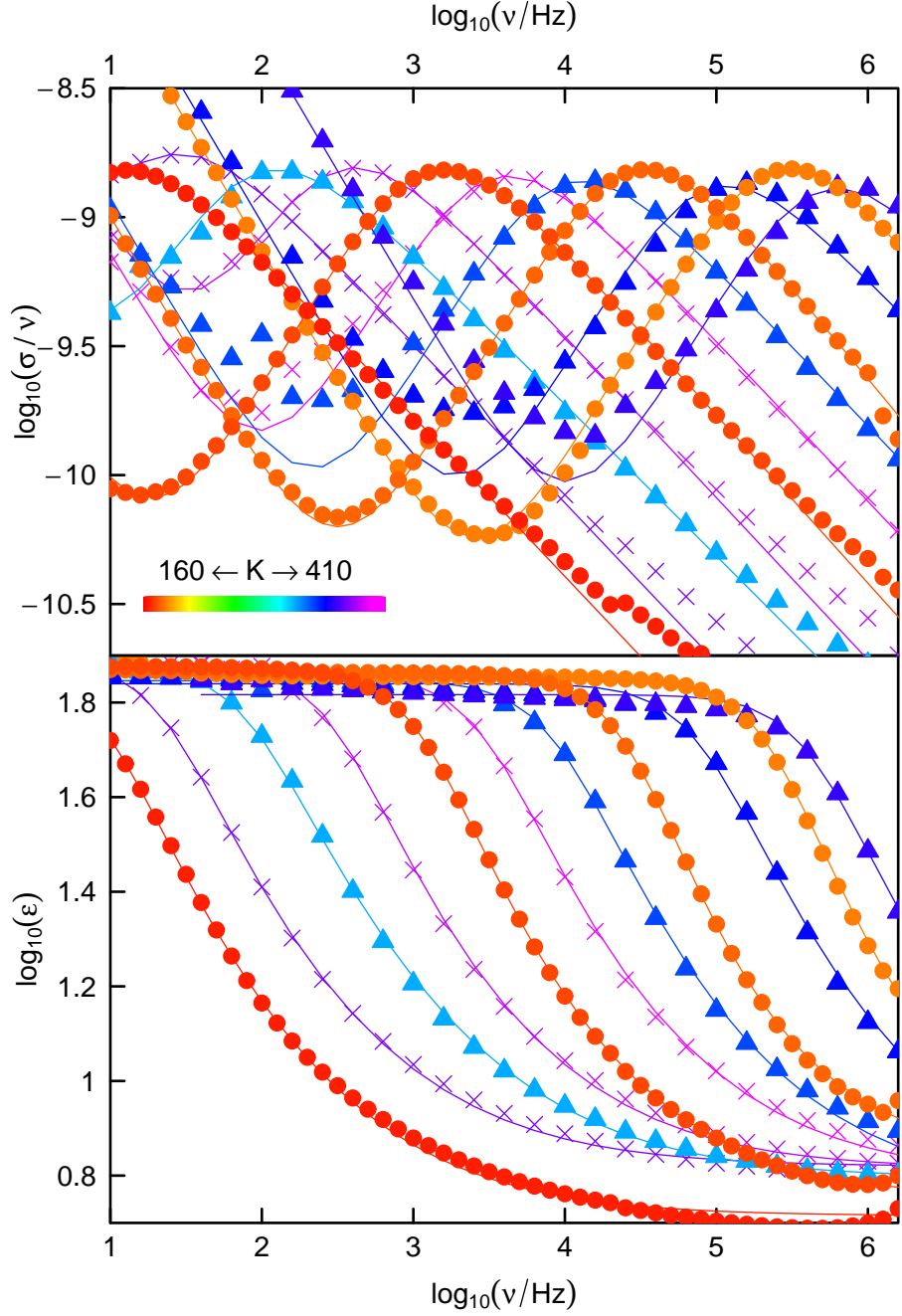}
\caption{Imaginary (upper panel) and real (lower panel) parts of dielectric permittivity measured along three isobars at different temperatures: $\bullet$ - 0.1 MPa (T=166.7, 173.8, 179.9, 185.4~K); $\blacktriangle$-2.3 GPa (T=326.0, 345.4, 358.5, 370.4~K); $\times$-3.3 GPa (T=384.3, 396.3, 406.1~K). In all cases, the temperature rise at fixed pressure shifts the dispersion feature toward higher frequencies. The color key encoding the temperature is also shown in the upper panel. Thin lines are the fits according to Eq.~(\ref{eq:cd}).}
\label{fig:ibar}

\end{figure}

\begin{figure}
  \includegraphics[width=0.8\columnwidth]{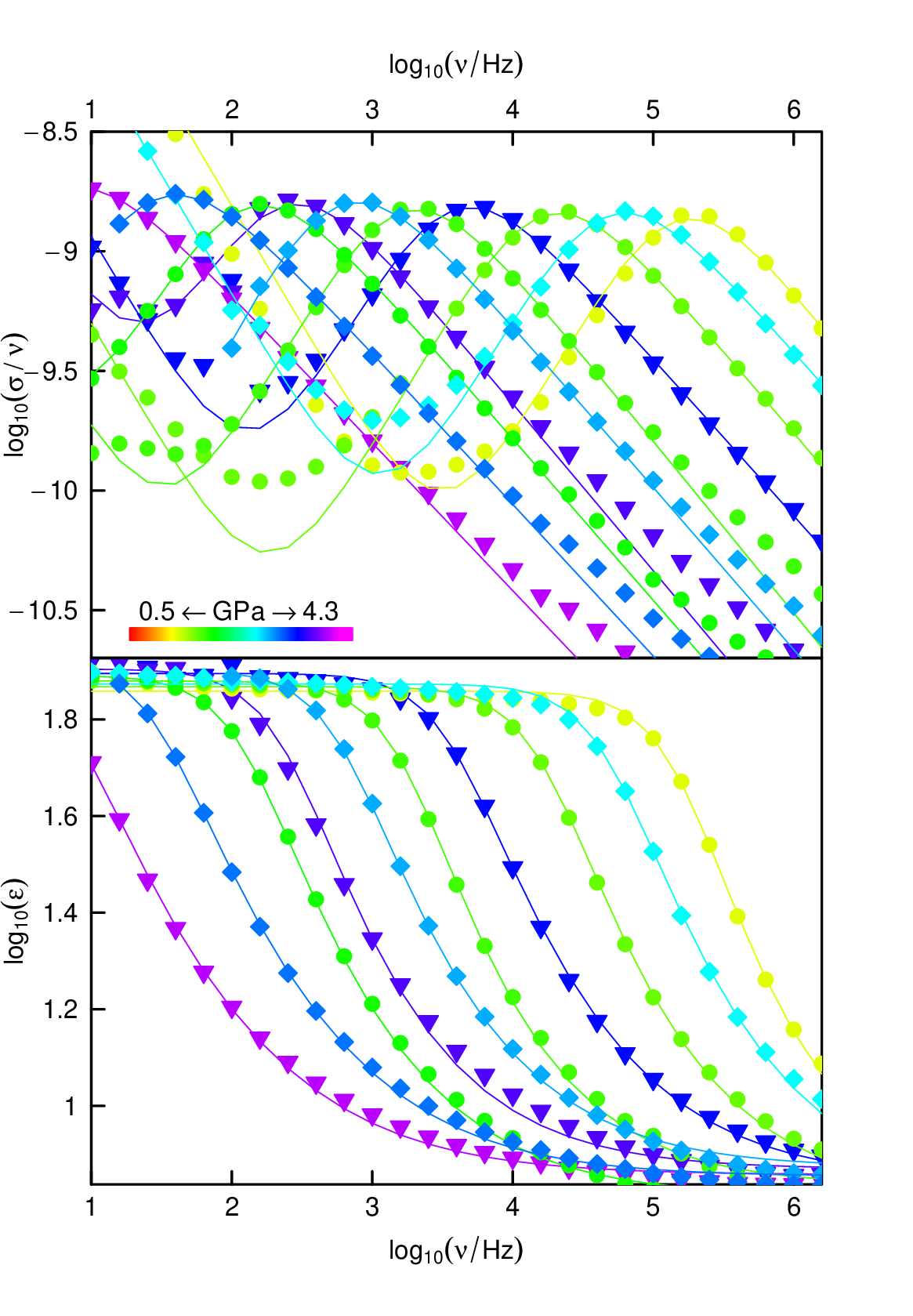}
\caption{Imaginary (upper panel) and real (lower panel) part of dielectric permittivity measured along three isotherms at different pressures: $\bullet$ - T=291.1~K (P=0.97, 1.38, 1.51, 1.65~GPa); $\blacktriangledown$-372.9~K (P=2.55, 2.86, 3.05~GPa ); $\blacklozenge$-410~K (P=3.41, 3.76, 4.1~GPa). In all cases, the pressure rise at fixed temperature shifts the dispersion feature toward lower frequencies. The color key encoding the pressure is also shown in the upper panel. Thin lines are the fits according to Eq.~(\ref{eq:cd}).}
\label{fig:itherm}
\end{figure}

\begin{figure}
  \includegraphics[width=0.8\columnwidth]{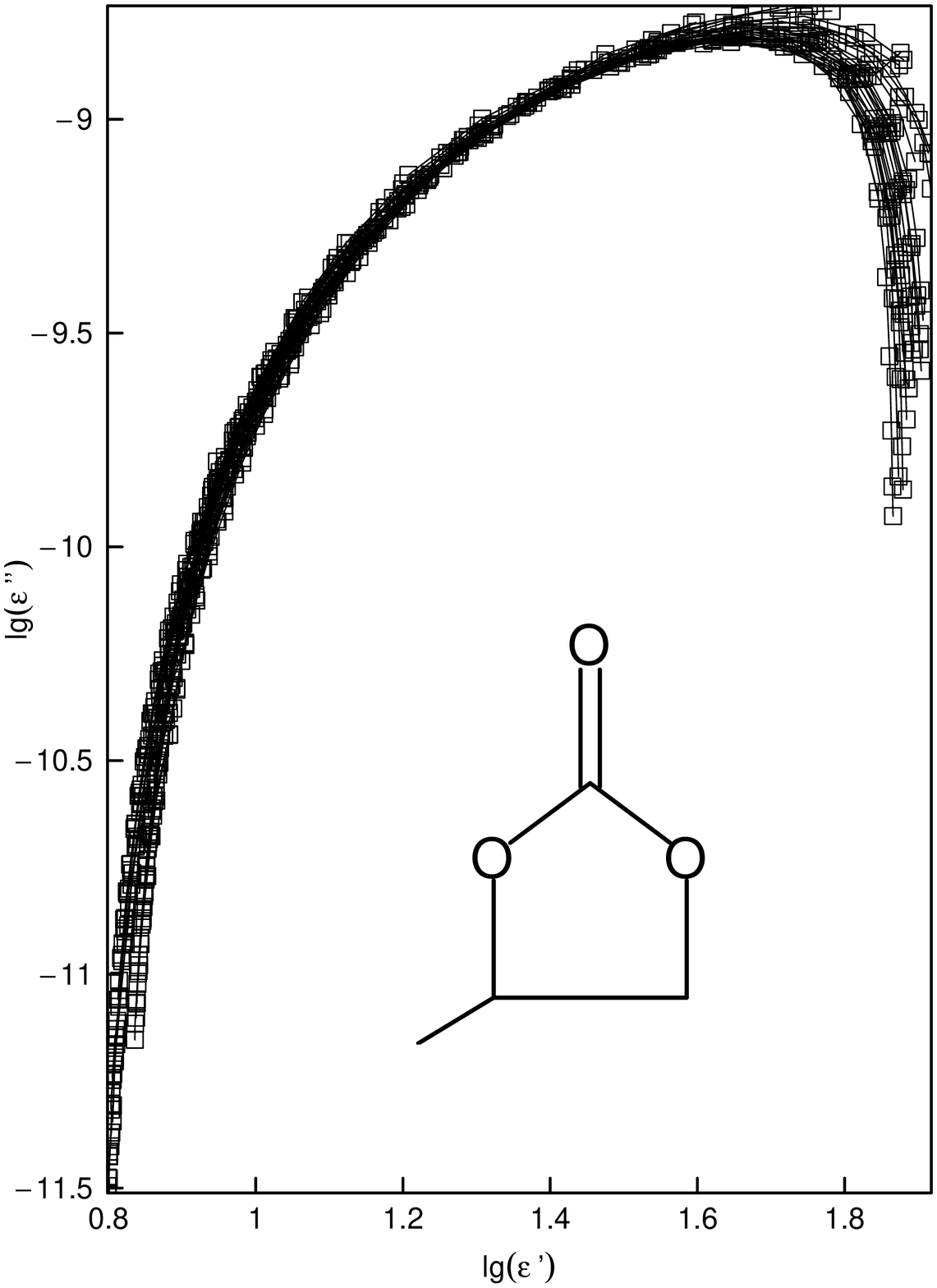}
\caption{
Cole-Cole plot of complex permittivity data taken in a single experiment in the range P=1.1 -- 4.1 GPa and T=292 -- 410 K. The PC molecule is shown schematically in the inset.}
\label{fig:cc}
\end{figure}

Some examples of raw dielectric spectroscopy data taken at several isobars and isotherms are presented in  Figs.~\ref{fig:ibar},\ref{fig:itherm} respectively. As it can be easily seen from these data, both the pressure rise and the temperature lowering shifts $\alpha$-relaxation to lower frequency range which eventually would lead to the liquid vitrification. Although this shift is accompanied with slight broadening of dispersion feature but in our temperature and pressure range we didn't observe any indication of $\beta$-relaxation. The absence of the secondary relaxation is easily demonstrated by the Cole-Cole plot of experimental data collected in a single ``isobar-isotherm'' run in P=0 -- 4.1 GPa and T=292 -- 410 K region presented in Fig.~\ref{fig:cc}. Despite the small variation in the dielectric permittivity in the low-frequency range (that is the right-hand part of Fig.~\ref{fig:cc}), high-frequency data follow the same master curve without any sign of a secondary process. In that respect, PC provides an  example of the glass-forming liquid where an universal behavior and single-mode relaxation are registered in the record-breaking range of density changes (almost twice).

Obviously, the stability of structural relaxation in PC under pressure rules out the possibility of irreversible polymerizations which occurs in many organic compounds with double bonds (such as benzene, acytelene or CO$_2$ \cite{brazhkin:pu06}) under pressure. Although this result might be interesting by itself, but in our case the pressure range could simply be insufficient for polymerizations to take place (typical values are about tens of GPa).

For processing the experimental data and parameterization of the structural relaxation dispersion, we used the well-known Cole-Davidson formula \cite{davidson:jcp51} Eq.~(\ref{eq:cd}) with the optional DC contribution $\sigma_o/\nu$ to imaginary (i.e. reactive) part $\varepsilon''$, which was taken into account when the experimental data required it. Both parts of the complex permittivity were fitted simultaneously with the target function to be minimized being the squared sum of residuals (i.e. it was induced by the norm on the complex plane). This sort of fitting was previously used for the description of dielectric relaxation in PC at ambient pressure \cite{schneider:pre98}.
\begin{equation}
\begin{split}
\varepsilon'&=\varepsilon_\infty+\varepsilon_1 Re\left(\frac{1}{(1+i\nu/\nu_0)^\beta}\right)\\ 
\varepsilon''&=\sigma_o/\nu+\varepsilon_2 Im\left(\frac{1}{(1+i\nu/\nu_0)^\beta}\right)
\end{split}
\label{eq:cd}
\end{equation}

It should be noted that usually the theoretical description of relaxation in glasses and supercooled liquids involves the notion of the transition from Debye-like relaxation (present in liquids at high enough temperatures) to the  so-called ``stretched-exponent'' relaxation described by the Kolraush-Wigner-Watts (KWW) correlation function of the form: 
\begin{equation*}
\Theta(t)\sim \exp\left(-\frac{t}{\tau_0}\right)^{\beta}
\end{equation*}
 However, it turns out that in case of PC a fit of experimental data in the frequency domain by the Fourier transform of the KWW-function obtained at ambient pressure is inferior to the result obtained by the Cole-Davidson function \cite{schneider:jncs98,schneider:pre98}. Interesting to note, that although the Cole-Davidson function was known for a long time, it was only recently (and it seems for the first time) when the authors of Ref.~\onlinecite{turton-wynne:jcp08,turton-wynne:jcp09} took notice that this function is the Fourier image of a quite simple correlation function (here the $\beta$ is the same as in Eq.~(\ref{eq:cd})):
\begin{equation*}
\Theta(t)\sim t^{\beta-1}\exp\left(-\frac{t}{\tau_0}\right)
\label{eq:debye}
\end{equation*}
Therefore, this is a Debye-like function with a strongly time-dependent prefactor (because in practice $0<\beta<1$) which in some way may reflect the more correlated nature of molecules motion in supercooled liquid compared with that in the normal one.

\begin{figure}
\includegraphics[width=0.8\columnwidth]{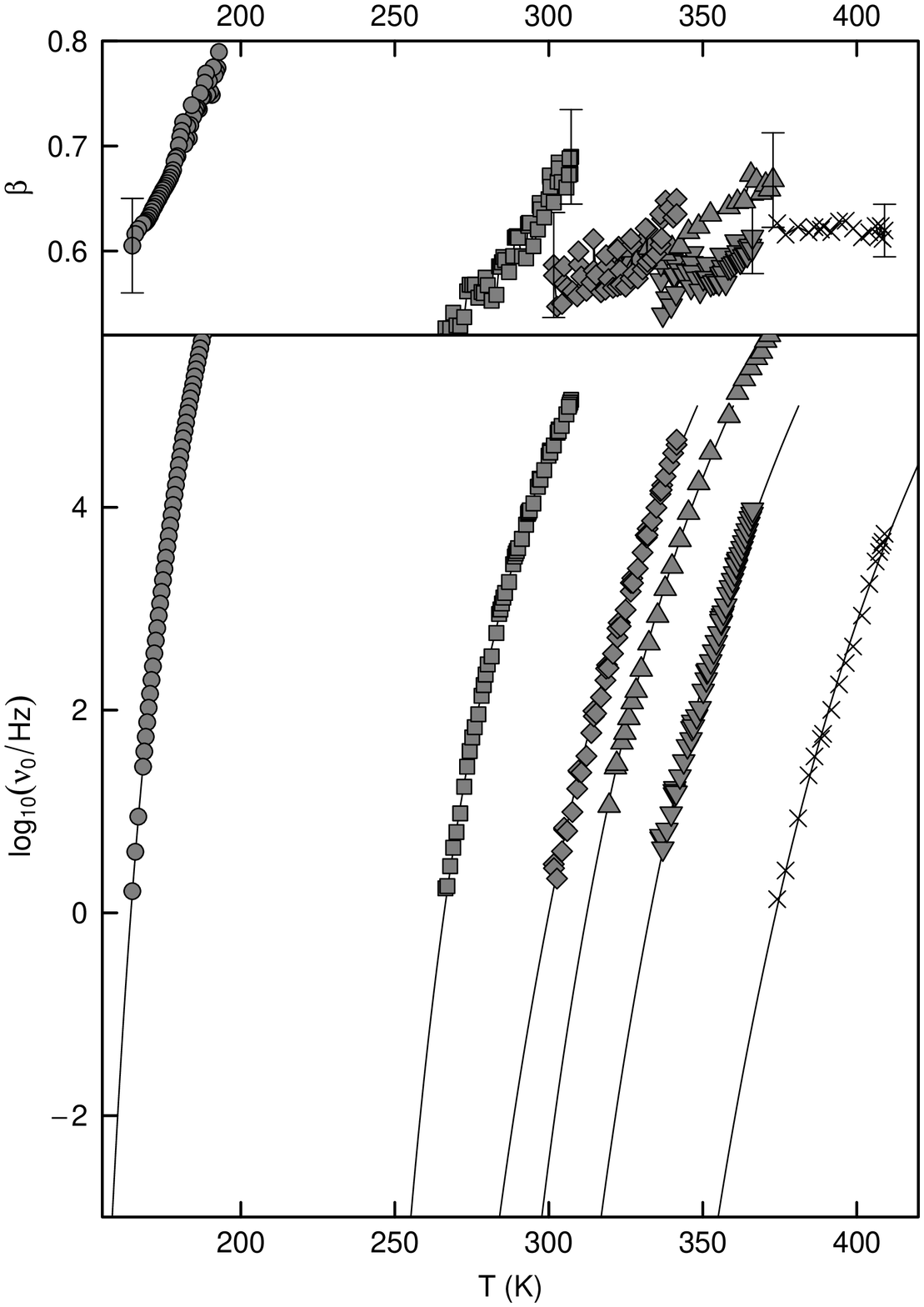}
\caption{ Variations in the  parameters $\beta$ (upper panel) and $\nu_0$ (lower panel) obtained by the fit of dielectric spectroscopy data with Eq.~(\ref{eq:cd}) along isobars. $\bullet$ -- P=0.1 MPa, $\blacksquare$ -- 1.5 GPa, $\blacklozenge$ -- 2.1 GPa, $\blacktriangle$ -- 2.3 GPa, $\blacktriangledown$ -- 2.7 GPa, $\times$ --3.3 GPa. Thin lines are the fits according to Eq.~(\ref{eq:vft}).}
\label{fig:vft}
\end{figure}

\begin{figure}
\includegraphics[width=0.8\columnwidth]{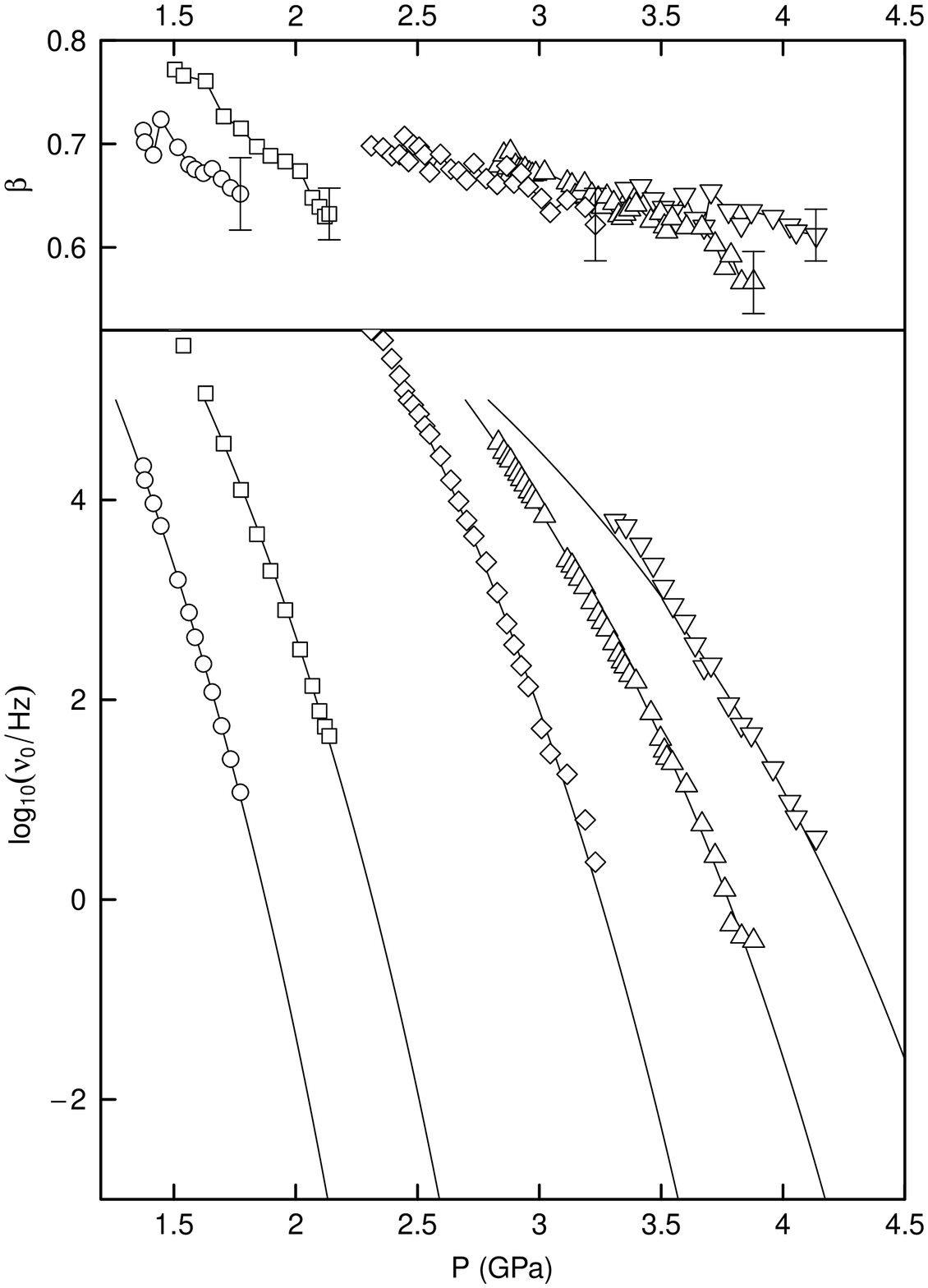}
\caption{Variations in the parameters $\beta$ (upper panel) and $\nu_0$ (lower panel)  obtained by the fit of dielectric spectroscopy data with Eq.~(\ref{eq:cd}) along isotherms. $\circ$ -- T=292.5 K, $\square$ -- 313.2 K, $\lozenge$ -- 373.1 K, $\triangle$ -- 394.1 K, $\triangledown$ -- 411.5 K. Thin lines are the fits according to Eq.~(\ref{eq:pvft}).}
\label{fig:pvft}
\end{figure}

The parameters $\beta$ and $\nu_0$ characterizing the structural relaxation of supercooled PC and  obtained by the fitting procedure with Eq.~(\ref{eq:cd}) are shown in Figs.~\ref{fig:vft}, \ref{fig:pvft} for isobaric and isothermic data respectively. For the quantification of the relaxation frequency $\nu_0$ we used  the well-known Vogel-Fulcher-Tamman (VFT) equation (Fig.~\ref{fig:vft}):
\begin{equation}
\log(\nu_0)=A+D_T\frac{T_0}{T-T_0}
\label{eq:vft}
\end{equation}
or its less known but still widely-used isothermic counterpart (PVFT) \cite{rzoska:jpcm08,bielowka:pre04}(Figs.~\ref{fig:pvft}):
\begin{equation}
\log(\nu_0)=A+D_p\frac{P}{P_0-P}
\label{eq:pvft}
\end{equation}

The accuracy of the exponent $\beta$ determination (which in some way can be regarded as a measure of dispersion width) according to Eqs.~\ref{eq:vft},\ref{eq:pvft} is not very high, but comparing the trend of parameter variations with its typical errors one may conclude that the dependence of this parameter on the external conditions (P or T) became weaker with the pressure rise. While the ambient pressure data suggest the temperature-dependent range 0.8 -- 0.6 (our estimations coincides with the results of Ref.~\onlinecite{schneider:pre98}), the high-pressure data point to the stabilization of this parameter around the constant (temperature- and pressure-independent) value $\beta=0.6 \pm 0.05$ (see Figs.~\ref{fig:vft},\ref{fig:pvft}).

Summarizing the dielectric spectroscopy data, we should once more stress  a remarkable stability of $\alpha$-relaxation in PC under high pressure conditions, which manifests itself in the ``topological'' invariance of the complex permittivity on the complex plane (Fig.~\ref{fig:cc}). One may even say that PC is the record-breaking molecular glassformer, in which the $\alpha$-relaxation stays intact up to the pressure 4.2 GPa while in other types of molecular glassformers much lower pressures cause the secondary relaxation to appear even, if it was not observed at ambient pressure (for typical examples see Ref.~\onlinecite{paluch:prl03}). It seems that t his fact can be somehow linked with the nature of intermolecular forces (for example, the presence or absence of hydrogen bonds between molecules) in the studied liquid.
\section{Ultrasonic measurements}
\begin{figure}
\includegraphics[width=0.8\columnwidth]{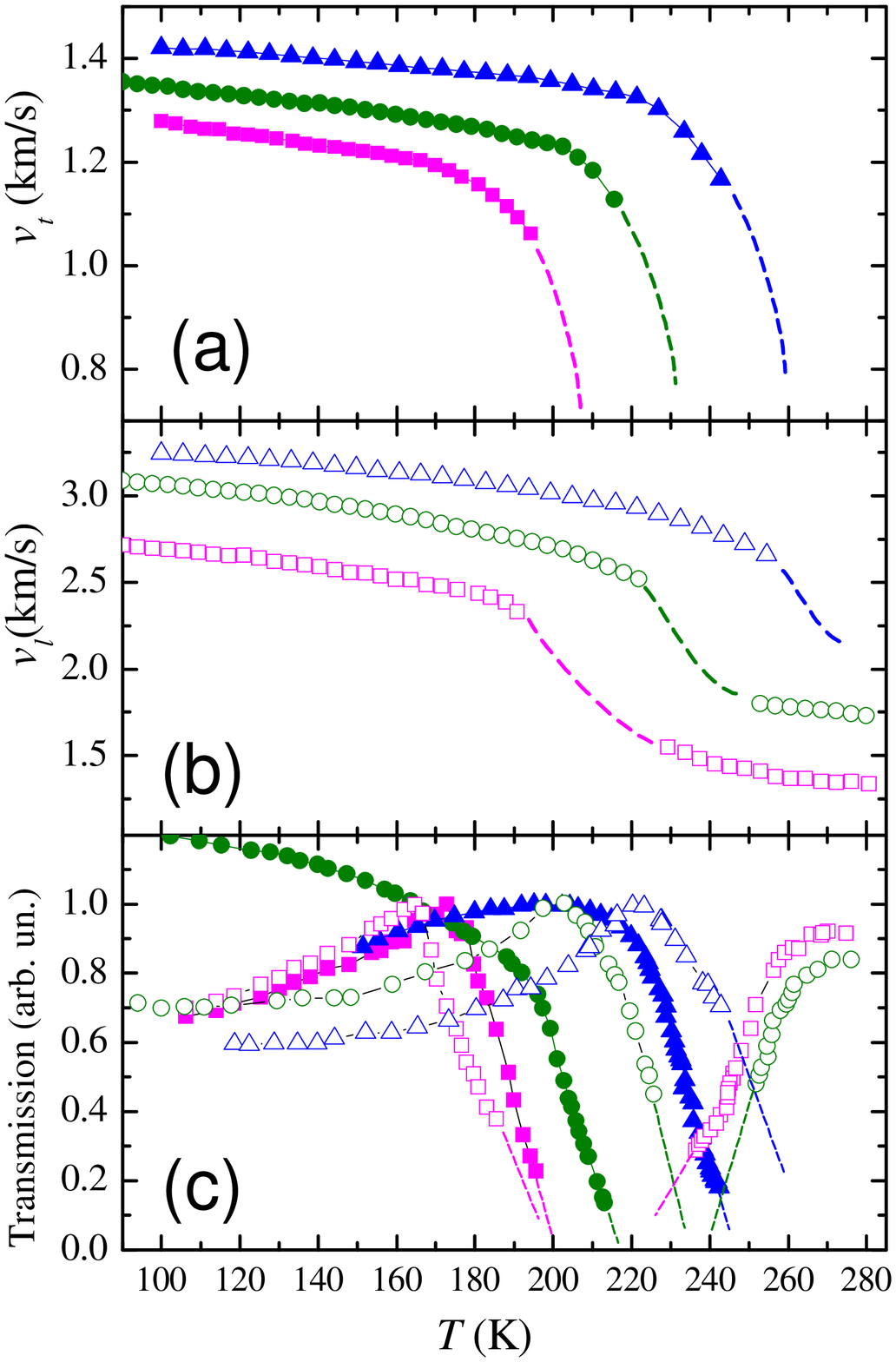}
\caption{Temperature dependences of the transverse $v_t$ (a) and longitudinal $v_l$ (b) ultrasound wave velocities and normalized transmission coefficients (c) measured at different pressures: $\square$ corresponds to $P=0.05$ GPa, $\circ$ - 0.55 GPa, and $\triangle$ - 0.9 GPa (solid symbols are for transverse waves, open -- for longitudinal one). Dashed lines are guide to eyes.}
\label{fig:us1}
\end{figure}

Fig.~\ref{fig:us1} presents the temperature variation in isobaric ultrasound wave velocities during the glass-to-liquid transition. Determination of the initial velocity and density values at ambient pressure will be discuss later. The normalized transmission (reverse attenuation) coefficients in Fig.~\ref{fig:us1}(c) were measured during the same experiments. The significant transmission drop evidently corresponds to the glass transition, when any accurate measurements are impossible (see Sec.~\ref{sec:meth}), and we can only interpolate the corresponding dependences above and below the glass transition. The dependence of longitudinal wave velocity (as well as transmission) at 0.9 GPa is incomplete for the liquid side since the capsule with PC leaks at this pressure. It is clear that the glass-to-liquid transition is accompanied by a lost of transverse ultrasound and a considerable drop of longitudinal wave velocity. The presented transmission data are insufficient for any decisive conclusions. Nevertheless, one can see a regular enhancement of the longitudinal wave transmission upon heating just before the glass transition. This effect seems to be related with the healing of defects (serving as scattering centers) in the glassy state as a precursor of intense diffusion of all atoms.

\begin{figure}
\includegraphics[width=0.8\columnwidth]{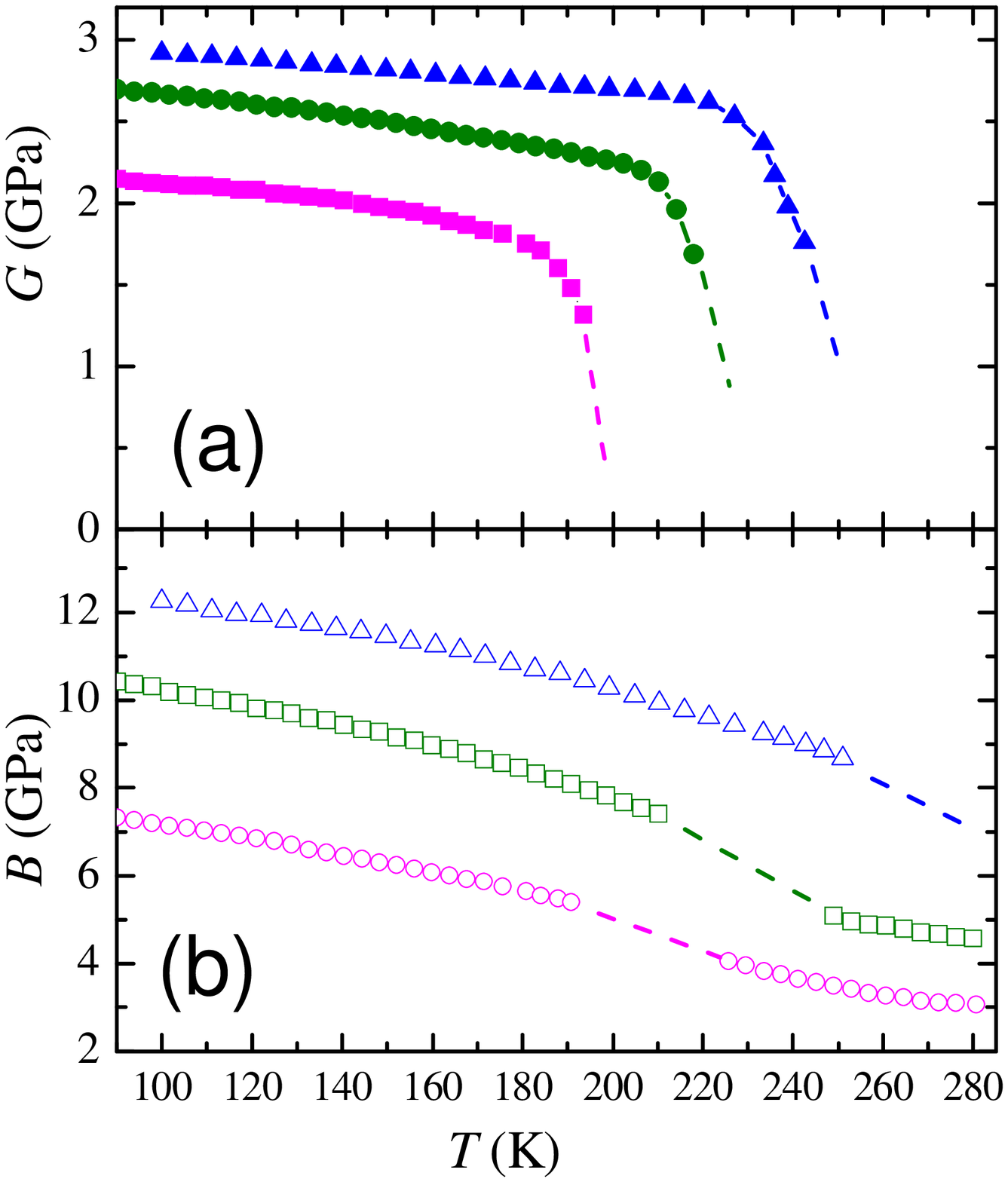}
\caption{ Temperature dependences of shear $G$ (a) and bulk $B$ (b) for the same pressures as in Fig.~\ref{fig:us1}, where the notations are also the same. Dashed lines are guide to eyes.}
\label{fig:us2}
\end{figure}

 The shear $G$ and bulk $B$ moduli of PC calculated according to Eqs.~(\ref{eq:B-G}) with the use of the data from Fig.~\ref{fig:us1} are presented in Fig.~\ref{fig:us2}. The shear modulus naturally demonstrates the same trends with temperature and pressure changes as transverse wave velocity in Fig.~\ref{fig:us1}(a) and the difference (although less spectacular than in Fig.~\ref{fig:us1}(b)) of the bulk moduli between the glassy and liquid states of PC is also visible. One should notice a considerable decrease in bulk modulus at a low pressure (0.05 GPa) from $\approx 7.5$ GPa in low-temperature glass to $\approx 3$ GPa in liquid PC near room temperature. Obviously, a similar decrease by a factor of $\approx 2.5$ should be observed at ambient pressure too.

\begin{figure}
\includegraphics[width=0.8\columnwidth]{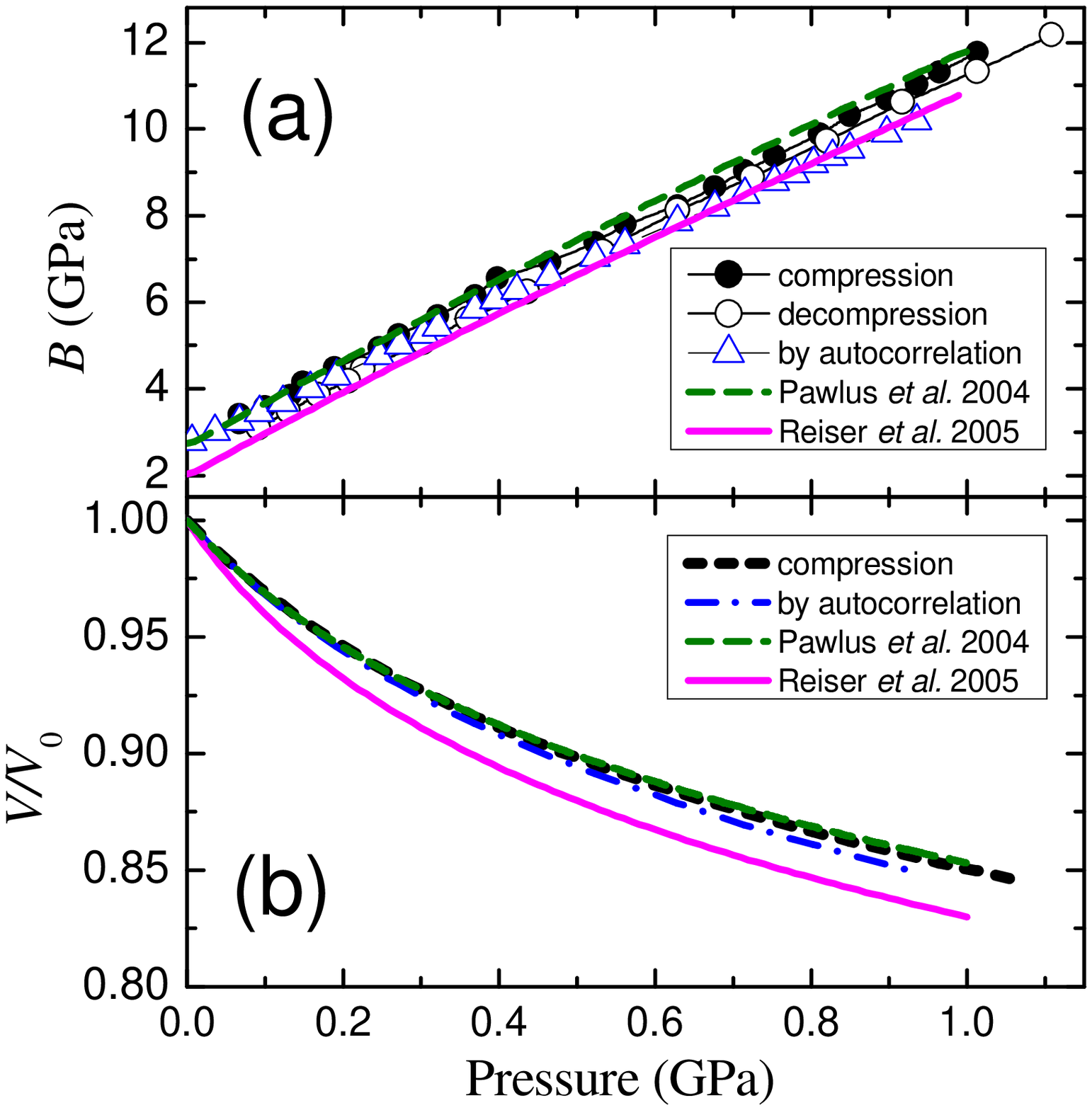}
\caption{ Pressure dependences of the bulk modulus $B$ (a) and isothermal equation of state (b) for the liquid PC at $T=295$ K, obtained from relative measurements of the transit time ($\bullet$ on (a) and the dashed line on (b)) and from autocorrelation measurements of the absolute transit time ($\triangle$ on (a) and the dash-dotted line on (b)). The bulk moduli calculated by the differentiation of the equations of state from Refs.~ \cite{reiser-kasper:prb05,pawlus:prb04} at $T=295$ K and the equations of state themselves  are also shown for comparison.}
\label{fig:us3}
\end{figure}

The bulk modulus of liquid PC can be directly determined from ultrasonic measurements of longitudinal wave velocity, since the shear rigidity of a liquid at ultrasonic frequencies is absent. The pressure dependences of bulk modulus and relative volume calculated from the isothermic ultrasonic data collected at $T=295$ K are presented in Fig.~\ref{fig:us3}. Here, our equation of states $V/V_0(P)$ was obtained by integration of the experimental pressure dependence of the inverse bulk modulus. The equations of state for PC from Refs.~\onlinecite{reiser-kasper:prb05,pawlus:prb04} are extrapolated up to 1 GPa and are shown for comparison, while the isothermal bulk modulus is calculated as $B=-V(\partial P/\partial V)_T$ from the analytic formulas $V/V_0(P,T=295$ K) published in these works. For our calculations we used the ambient pressure density of PC\cite{pawlus:prb04}, $\rho(295 {\rm K})\approx 1.19$ g/cm$^3$, and the longitudinal wave velocity from Brillouin measurements \cite{elmroth:prl92}, $v_l(295 {\rm K})\approx 1510$ m/s. Our data quantitatively agree  with the results of direct volumetric measurements \cite{reiser-kasper:prb05,pawlus:prb04}, but we should take into account that the equation $V/V_0(P,T)$ from Ref.~\onlinecite{pawlus:prb04} was developed for relatively low pressures (see comment in Ref.~ \onlinecite{reiser-kasper:prb05}), while the equation from Ref.~\onlinecite{reiser-kasper:prb05} was tested only at pressures up to 0.7 GPa. Our data does not allow us to tell which one is best for ambient conditions. The better correspondence between the initial value of the bulk modulus in our dependence and that from Ref.~\cite{pawlus:prb04} may be influenced by the choice of the initial longitudinal wave velocity from Ref.~\cite{elmroth:prl92}. Possibly, the equation of state from Ref.~\onlinecite{reiser-kasper:prb05} covers much wider intervals of pressure and temperature, but yields a less accurate value of the bulk modulus at the particular point (1 bar, 295 K). At the same time, our autocorrelation measurement in Fig.~\ref{fig:us3} seems to be more accurate (it gives the absolute transit time of the sound wave), and the corresponding pressure dependence $B(P)$ becomes closer to the data from Ref.~\onlinecite{reiser-kasper:prb05} with pressure increase. 

As one can see from Fig.~\ref{fig:us3}(a) the pressure dependence  $B(P)$ obtained from the autocorrelation data is almost linear in the studied pressure ranges and its pressure derivative  $B'(P) \approx 8$ is characteristic of  the Lennard-Jones system (see, e.g. , Ref. \cite{br:pma02}). Indeed, for the $m-n$ potential Eq.~(\ref{eq:lj}), it can be easily demonstrated that \cite{alexandrov:jetp87}:
\begin{equation}
B'(P)=(m+n)/3+2
\label{eq:mn}
\end{equation}

\begin{figure}
\includegraphics[width=0.8\columnwidth]{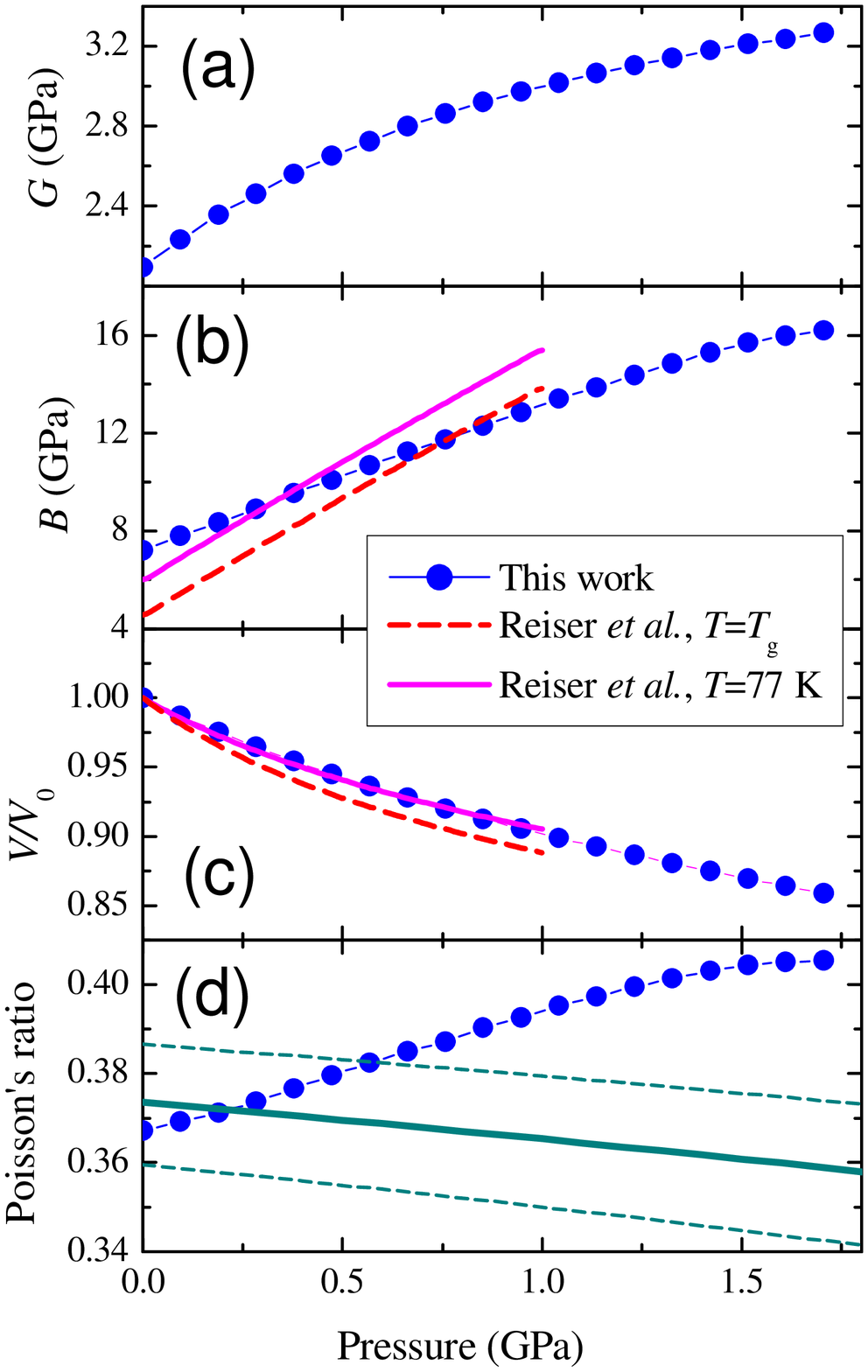}
\caption{ Pressure dependences of the shear $G$ (a) and bulk $B$ (b) moduli, the equation of state (c) and Poisson's ratio (d) for glassy PC at 77 K ($\bullet$). The equations of state from Ref. \cite{reiser-kasper:prb05} for $T=77$ K and $T= T_{\rm g}$ (156 K at 1 bar \cite{reiser-kasper:prb05}) and the pressure dependence of bulk modulus calculated from them are also shown as lines on panels (b) and (c). The solid and dashed lines in panel (d) correspond to the correlation between Poisson's ratio and the fragility index discussed in Sec.~\ref{sec:conclusion}.}
\label{fig:us4}
\end{figure}

The elastic moduli for glassy PC measured at 77 K (see Fig.~\ref{fig:us4} (a-b)) are obtained from the ultrasonic measurements of both transverse and longitudinal wave velocities. The values of the elastic moduli have a larger uncertainty (about 5-10\%) compared to the data taken at ambient temperature. This uncertainty is induced  to a large extent  by the uncertainty in the choice of the initial density value  $\rho (77 {\rm K})$ and absolute sound velocities of glassy PC. The extrapolation of the liquid equation of state from Refs.~\onlinecite{reiser-kasper:prb05,pawlus:prb04} yields an overestimated density value $\approx 1.4$ g/cm$^3$. The corrected value of $\rho(77$ K) for PC can be obtained by the comparison of equations of state from Refs.~\onlinecite{reiser-kasper:prb05,pawlus:prb04} with the temperature dependence of $\rho$ and the thermal expansion coefficient below $T_{\rm g}$ for a relatively close glassformer, glycerol \cite{blazhnov:jcp04}, and subsequent accounting for the corresponding shift of $T_{\rm g}$. We obtained $\rho (77 {\rm K})\approx$1.34--1.35 g/cm$^3$ for PC. This value is in accordance with the results of the molecular dynamic simulation \cite{eckstein:jcp00}. The longitudinal wave velocity at ambient pressure can be extrapolated down to 77 K from the Brillouin scattering study of PC \cite{elmroth:prl92}, but similar procedure is obviously impossible for the transverse velocity. So, for calculation of the transverse velocity, knowing the absolute transit time of the sound wave in the sample from autocorrelation measurements, we approximated the height of sample (the sound wave path) in two ways: first, from longitude measurements as mentioned above, and, second, from the height of the liquid PC sample at 295 K scaled by the thermal expansion of PC, $\rho(295 {\rm K})/ \rho(77 {\rm K})$. The contribution of the thermal expansion of the capsule in the last case was negligible but could be taken into account by the use of the known tabular data for materials of the capsule. Our data in Fig.~\ref{fig:us4} are in qualitative and partially in quantitative accordance with the equation of state from Ref.~\onlinecite{reiser-kasper:prb05} extrapolated to $T_{\rm g}$ and 77 K. We did not find in the literature the sound velocities $v_l$ and $v_t$ for glassy PC at 77 K, but these data are available\cite{torchinsky:jcp09} at $T_{\rm g}=$159.5 K for the frequency $\approx 600$ MHz, $v_l=2608$ m/s and $v_t=1031$ m/s, and they are quite close to the values obtained by us at 77 K and ambient pressure: $v_l=2710$ m/s and $v_t=1250$ m/s. We can also make comparison with glycerol which has similar values of density, $T_{\rm g}$ and molecular weight $M$ ($M=92$ for glycerol C$_3$H$_8$O$_3$ and $M=102$ for PC C$_4$H$_6$O$_3$),  but  appreciably different magnitudes of sound velocities at 77 K \cite{ramos:pm04} -- $v_l=3720$ m/s and $v_t=1890$ m/s. This difference only partially could be accounted for by the difference of values of $T_{\rm g}$ in these two glassformers. In our opinion, a more sound explanation should be probably based upon taking into account a qualitative difference of intermolecular forces in these glasses. In contrast to PC, the molecule of glycerol has three (OH)$^-$ groups, so hydrogen bonding between molecules in glycerol enhances the elasticity response. On this ground we can predict a relative decrease in  the elasticity difference between PC and glycerol under even higher pressures.

The striking difference in the behaviors of the bulk moduli of glassy (Fig.~\ref{fig:us4}(b)) and liquid PC (Fig.~\ref{fig:us3}(a)) is a  more pronounced nonlinear pressure dependence and higher values of the bulk modulus  in the glassy state. A similar relation between the glass and liquid bulk moduli are evidently follow from the Brillouin scattering data on the longitudinal wave velocities \cite{elmroth:prl92} too.  The observed pressure derivative $B'(P)$ for PC glass decreases from $\approx 6.5$ (at ambient pressure) down to $\approx 5$ (at $P=$1.7 GPa), and these values are considerably lower  than those for liquid PC. The observed values of $B'(P)$ can be interpreted formally as an crossover taking place in PC from  the Lennard-Jones Eq.~(\ref{eq:lj}) to soft-sphere model Eq.~(\ref{eq:ss}) with repulsion exponent $n=12$. In the last case, Eq.~(\ref{eq:mn}) is modified to a simpler equation $B'(P)=n/3+1$. So, with pressure increase PC glass tends to the soft-sphere system with $B'(P)=5$ ($n=12$). This conclusion is compatible with a more general  observation that the compression of the Lennard-Jones systems by factor of 10--15 \% results in the $B'(P)$ decrease down to 6.5--7 \cite{br:jetpl01}. Nevertheless, we can not consider the glassy PC as a system with purely central intermolecular potential (e.g., like the Lennard-Jones or soft-sphere system), because of the high Poisson coefficient value $\sigma$ observed in glassy PC which is equal to $\approx 0.37$  at ambient pressure and increases with pressure (Fig.~\ref{fig:us4}(d)).  Comparison of Poisson's ratio with other relaxation properties of PC will be discussed in more details in the next section, but here we should note that the Poisson coefficient values typical of metals (0.3--0.4)\cite{br:pma02} were previously observed in the  molecular solids, e.g., fullerites C$_{60}$ and C$_{70}$ \cite{yag:jpcs10} too.

\section{Discussion and conclusive remarks}
\label{sec:conclusion}
Due to the limited frequency range, the glassification temperatures and pressures can be inferred from the dielectric spectroscopy data by the  extrapolation of isobaric and isothermic data according to VFT Eq.~(\ref{eq:vft}) or PVFT Eq.~(\ref{eq:pvft}) relations, respectively. Here we mean by glassification the conditions when the liquid is in the ``practically'' glassy state, in other words, when the $\nu_0=10^{-3}$~Hz.  The exact value is very much arbitrary but the thermodynamic parameters defined in such a way usually match quite well with the ones obtained by the calorimetric methods \cite{wang-velikov-angell:jcp02}. This fact is not surprising taking into account the strong variation of the VFT function in vicinity vicinity of its root so the variation in the exact value of relaxation time or the method for its determination  is not crucial. 

\begin{figure}
\includegraphics[width=0.7\columnwidth]{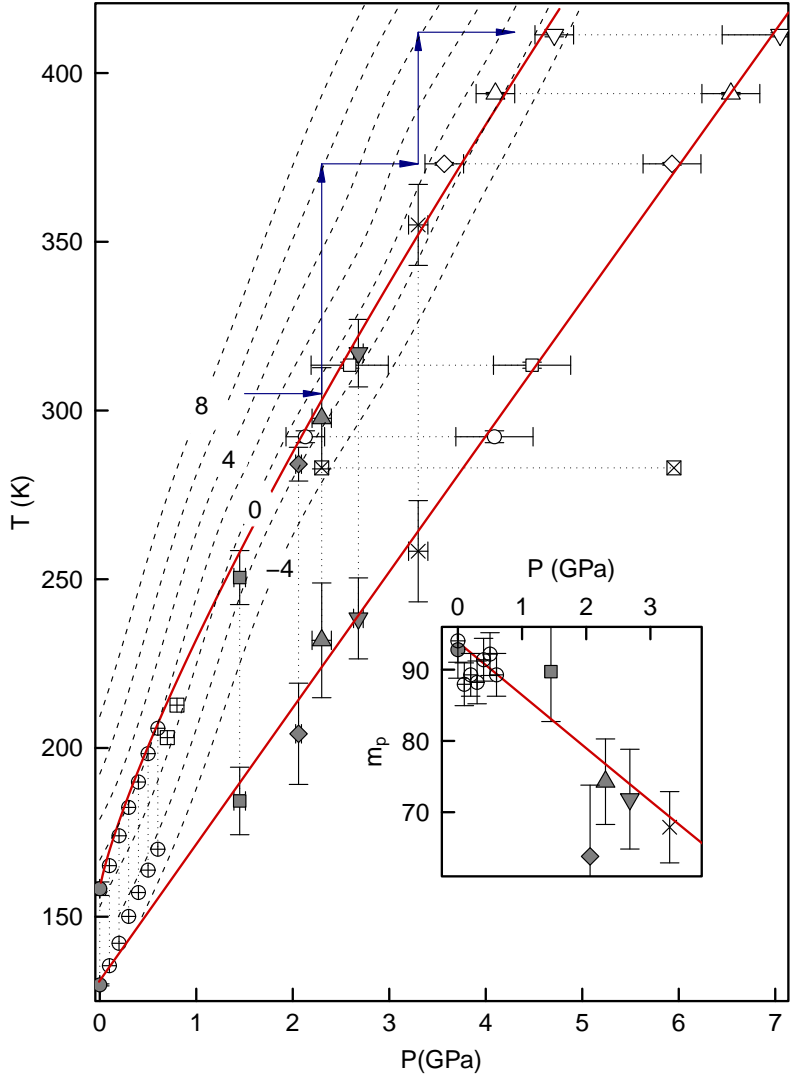}
\caption{Evolution of the characteristic frequency of structural relaxation in PC on P-T table. Symbols on the graphic mark the values of glassification parameters for ``practical'' (upper curve) and ideal (lower curve) glass based on the VFT (Eq.~\ref{eq:vft}) or PVFT (Eq.~\ref{eq:pvft}) extrapolations while the directions of dotted lines indicate which one of the approximations was used. The dashed lines mark the $lg(\nu_{0})$ levels obtained by the interpolation of dialectic spectroscopy data by the ``thin-plate'' spine. The pressure evolution of the isobaric fragility index $m_p$ is shown in the inset. The symbol shapes on graphics correspond to the symbols in Figs.~\ref{fig:vft}, \ref{fig:pvft}. The only exception is the literature data: isobaric ones $\oplus$  -- from Ref.~\cite{reiser-kasper:epl06}, $\boxplus$ -- from Ref.~\cite{rzoska:jpcm08} and isothermic data $\boxtimes$ -- from Ref.~\cite{pawlus:prb04}. The upper thick solid curve in the main panel is the fit of $T_g$ with Eq.~(\ref{eq:tg}), while the other solid lines on the main plot and in the inset are just guide for eyes. Thin arrows indicate the trajectory of P-T parameter variations in the single experiment.}
\label{fig:p-t}
\end{figure}

It is interesting to compare these parameters with the parameters $T_0$ and $P_0$ from the Eqs.~(\ref{eq:pvft}),~(\ref{eq:vft})  when $\nu_0$ tends to zero (that is the relaxation time diverges), which usually labeled as ``ideal-glass'' state.  Although the exact physical sense of ``ideal-glass'' parameters (as well as a theoretical ground of the VFT equation) is still a subject of debates \cite{mckenna:np08,hecksher:np08}, but it seems to us worth plotting both types of the parameters ($T_g$, $P_g$ on the one hand and $T_0$, $P_0$ on the other) in the same figure (Fig.~\ref{fig:p-t}) where we also add some available literature data. 

The first straightforward conclusion one can draw from this graphic is a slight upward curvature  of glassification temperature $T_g$ dependence on pressure, which is a good indication that the soft-sphere model is well applicable to the description of PC structural relaxation in  wide pressure range. Because one may consider $T_g$-curve on P-T plane as isochronous curve where the relaxation time is constant, so one may naturally expect a scaling relation for it which is a consequence of Eq.~(\ref{eq:scaling}) \cite{hoover:jcp71,en*stishov:up74,en*zhakhovsky:zhetp94,cape:jcp80}:  
\begin{equation}
T_g \sim P^{n/(n+3)}
\label{eq:tg}
\end{equation}
The fitting of the experimental dielectric spectroscopy data for T$_g$ (see Fig.~\ref{fig:p-t}) with the formula Eq.~(\ref{eq:tg}) yields the exponent value $n=12.9$ close to the value obtained earlier from the viscosity data \cite{casalini:jcp08}. 

From Fig.~\ref{fig:p-t} one can draw one more conclusion regarding the P-T evolution of ``ideal-glass'' parameters. Although $P_0$ and $T_0$ values are defined implicitly by two phenomenological equations, both follow the same curve on the P-T plane, which may indicate that both equations  (\ref{eq:vft}) and (\ref{eq:pvft}) (especially the latter one as it is less widely accepted than the former) describe well the  vitrification process taking place in supercooled PC. Another non-parametric approximation of experimental data with the ``thin-plate'' spline \cite{wood:06,R:fields} shown in  Fig.~\ref{fig:p-t} by dashed line provides an independent validation of both these equations. 

On the inset of Fig.~\ref{fig:p-t} we present the widely used isobaric fragility index $m_p$ defined as:
\begin{equation*}
m_p=\frac{1}{T_g} \frac{\partial \log_{10}(\nu_0)}{\partial 1/T}|_{T=T_g}
\end{equation*}
which quantifies how far away is the temperature dependence of structural relaxation in liquid from the Arrhenius one.

This parameter can be measured directly by calorimetric methods at the glassification point  \cite{wang-velikov-angell:jcp02}, but in our case we can only infer this value from parameters of the VFT fit according to formula:
\begin{equation}
m_p=\frac{D_TT_0T_g}{\ln(10)(T_g-T_0)^2}
\end{equation}
In other words, the fragility index is a composite parameter evaluated through a combination of all of the free parameters present in VFT equation (parameter $A$ from Eq.~(\ref{eq:vft}) is implicitly defined by the selection of $T_g$).

Although the widely cited value at ambient pressure $m_p=104$ (based on dielectric spectroscopy data from Ref.~\onlinecite{huck:je82}) places PC as a very fragile glassformer, later measurements yield  slightly smaller values: $m_p=99$ from calorimetric measurements \cite{wang-velikov-angell:jcp02} and $m_p=94 \pm 3$ from dielectric ones \cite{reiser-kasper:epl06,schneider:pre98}. The value obtained in this work at ambient pressure practically coincides with the last one ($m_p=92 \pm 5$) but  it seems to us that the limited pressure range does not allow the authors of Ref.~\onlinecite{reiser-kasper:epl06} to recognize the downward trend which is easily discernible when their data are plotted in the wider pressure range (their results are presented in inset Fig.~\ref{fig:p-t}). Still the fragility index variation of 7\% per GPa found by us is considerably lower than the trend suggested by the authors of  Ref.~\onlinecite{casalini-roland:prb05}. The weak pressure dependence of fragility may be related to the results obtained from computer simulations for soft-sphere model \cite{michele:jpcm04} which predicted that the substances described by the power-law  repulsive potential should have a high value of the fragility invariant from the exponent value.

Ultrasound measurements reveal a significant difference between the behaviors of the bulk modulus under pressure in the glassy and liquid states of PC. Although it was known  earlier \cite{elmroth:prl92}, the new data on a relatively simple glassformer PC allow us to confirm a great difference between the bulk modulus pressure derivatives in liquid (7 --  8.5) and glassy (4.5 -- 5.5) states. The value of $B'(P)$ directly depends on the type of intermolecular forces. The value of $B'(P)\approx$ 7 -- 8 eventually allows one to suppose the intermolecular potential of the Lennard-Jones type. On the other hand, significantly lower values registered in the glassy state $B'(P) \approx $ 5 -- 6 accompanied with quite high values of Poisson's coefficient $\sigma  \approx$ 0.35 -- 0.4  suggest a more complicated character of the intermolecular interaction. Obviously, the temperature lowering and/or pressure rising bring molecules in a liquid closer together so the molecules cannot be considered as quasi-spherical particles and more subtle details (like the deviation of molecules and their interaction from spherical symmetry) should be taken into account. On the other hand, the absence of $\beta$-relaxation in the PC under high pressure indicates that it is indeed one of the most experimentally simple model objects. In the future it will be desirable to carry out dielectric spectroscopy measurements of PC in the ultra-low frequency range $\nu=$10 -- 10$^{-3}$ Hz under high pressures  to demonstrate that in the high viscosity regime (which accompanies the glass-liquid transition) a crossover from one scaling type (with the exponent $n=$12 -- 13) to another  ($n=$8 -- 10) would actually take place.

We should mention, however, that the non-central character of intermolecular interactions in PC glass does not necessarily mean that scaling relations like Eqs.~(\ref{eq:scaling}),(\ref{eq:tg}) are not applicable in this case. Just the opposite -- we have already demonstrated that the glassification curve in PC does obey Eq.~(\ref{eq:tg}). There is no contradiction because the scaling relations are a consequence of the Klein theorem valid for a wide class of homogeneous potentials of which Eq.~(\ref{eq:ss})  is one special case. Nonetheless one can imagine a homogeneous but non-central potential, e. g. the one which includes all positional coordinates with the same power but different weight. There is no ``spheres'' in such a model but it is still ``soft''.

It is pertinent to consider the correlation\cite{novikov:n04} between the fragility index and Poisson's ratio $\sigma$ (or $B/G$ ratio for a homogeneous isotropic solid according to Eq.~(\ref{eq:sigma})) of different substances at ambient pressure, which was initially proposed for  all glassforming liquids, but after subsequent discussion\cite{johari-spyros:n06,novikov:prb06,johari:pm06} it was restricted to the class of non-metallic molecular glassformers\cite{sokolov:pm07} where PC surely belongs to. For nonmetallic glasses this correlation is described by the equation \cite{novikov:prb06}:
\begin{equation}
m_p=(29\pm 2)B/G - (12\pm 5).
\label{eq:mpsigma}
\end{equation}
It is interesting to note that if we look into original paper\cite{novikov:n04}, we will see that PC is an outlier due to its overestimated fragility index value ($\approx$ 104). The lower value about 90 brings this point closer to the main dependence.

Using the linear approximation for $m_p(P)$ from the inset of Fig.~\ref{fig:p-t} we plotted the expected from Eq.~(\ref{eq:mpsigma}) pressure dependence $\sigma(P)$ with maximal deviations in Fig.~\ref{fig:us4} (d). Whereas at ambient pressure $m_p$ and $\sigma$ correlate quite well, the experimental and expected $\sigma(P)$ dependences obviously deviate from each other. At 1 GPa this deviation is observed, even if we would suppose that $m_p(P)$ is pressure independent (according to Ref.~\onlinecite{reiser-kasper:epl06}). The correlation between $m_p$ and $\sigma$ (or $B/G$) for a fragile liquid, like PC, at ambient pressure can be described in a simplified manner in terms of a very heterogeneous potential landscape \cite{stillinger:n01}, when potential energy barriers and corresponding energies are wide spread. On the one hand, this means the fragile behavior during the glass transition and large value of $m_p$. On the other hand, low-barrier paths in the potential landscape provide additional possibilities for structural relaxation of glass against shear deformation, since it does not change the volume. For this reason, Poisson's ratio and the $B/G$ ratio in the fragile glass are higher. The $m_p$ decrease in PC with pressure indicates that the system becomes stronger, and low-barrier features of the potential landscape are forced away from the area achievable by the system. Although the $\sigma$ increase with pressure is not an absolute rule (see, e.g., $\sigma(P)$ for C$_{70}$ \cite{yag:jpcs10}), such a behavior is very typical for majority of amorphous and crystalline solids. Indeed, the greater relative increase in the  bulk modulus with respect to the shear one is naturally related to a simple observation that central repulsive forces (strongly increasing with pressure) are, as a rule, more important for compressibility than for the shear elastic response. Thus, it is clear that for PC different features of the potential energy landscape are responsible for the evolution of dielectric (basically rotational) response and elastic properties (basically translational). So we believe that the Eq.~(\ref{eq:mpsigma}) requires some modification in the higher pressure limit.

\begin{figure}
\includegraphics[width=0.8\columnwidth]{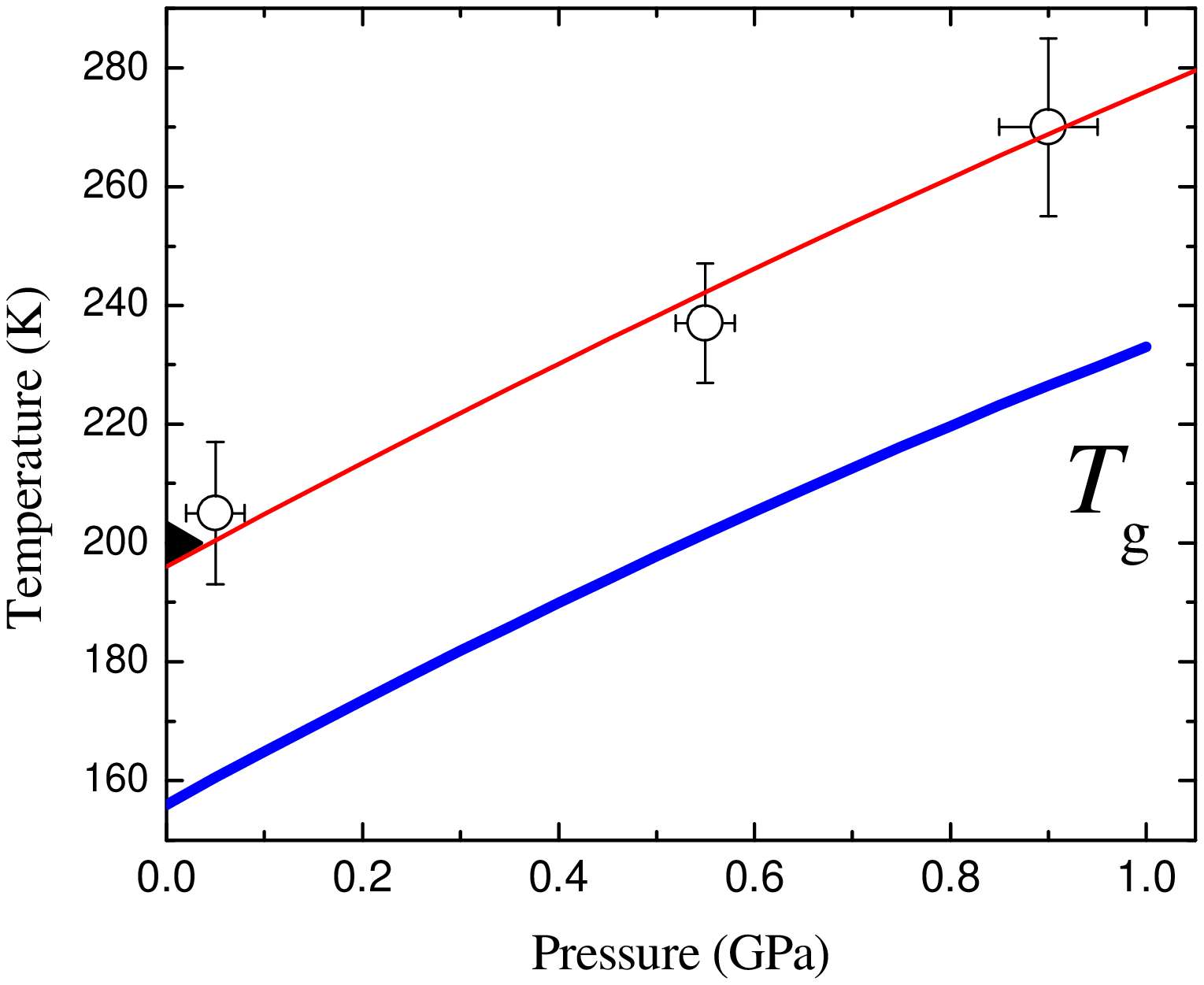}
\caption{Pressure-temperature points (open circles) corresponding to the minimum of the transmission of longitudinal wave. The horizontal bars correspond to the uncertainty of pressure measurements and the vertical bars show entire possible interval (see Fig.~\ref{fig:us1}). The thin line corresponds to the position of maximum of the imaginary part of dielectric permittivity at 10 MHz, and it is obtained by the extrapolation of the presented data. The black triangle shows the same at ambient pressure and is taken from Ref. \cite{schneider:pre98}. The pressure dependence of the glass transition temperature is taken from 
Ref.~\cite{reiser-kasper:prb05}.}
\label{fig:us5}
\end{figure}

One can compare the relaxation properties during the glass transition in the dielectric and elastic responses of PC. In Fig.~\ref{fig:us5} we plotted the points of maximum attenuation of longitudinal ultrasonic waves at 10 MHz and the position of maximum of the imaginary part of dielectric permittivity at 10 MHz from the literature at ambient pressure and by extrapolations of the presented data under pressure. There is good coincidence between the relaxation data from the dielectric spectroscopy and ultrasonic measurements. This is not surprising because 10 MHz is still a too high frequency for the glass transition, and characteristic temperatures are too far from $T_{\rm g}$ for phenomena like the heterogeneity at the glass transition \cite{chang:jpcb97} and the translation-rotation paradox for diffusion \cite{hall:prl97} to occur. Nonetheless application of ultrasonic technique for studying the elasticity of other glassformers combined with the dielectric spectroscopy data may be very useful for understanding how different features of the energy landscape affect the evolution of the glass relaxation dynamics under pressure.

\begin{acknowledgements}
This work was supported by the Russian Foundation for Basic Research (Russian-German Joint Program No. 09-02-91351 and Grants Nos. 10-02-01407 and 11-02-00303), by the Programs of the Presidium of the Russian Academy of Sciences, and by the Deutsche
Forschungsgemeinschaft via the German-Russian joint research project, Grant No. 436 RUS113/992/0. 

We are also grateful to A.V. Rudnev for technical assistance in the accomplishment  of low-temperature measurements under pressure, Profs. P. Lunkenheimer and A. Loidl for fruitful discussions, S.M. Stishov for support and interest in our work and Prof. Anthony N. Papathanassiou for very pertinent remarks.
\end{acknowledgements}
%

\end{document}